# THE ATMOSPHERIC MUTUAL COHERENCE FUNCTION FROM THE FIRST AND SECOND RYTOV APPROXIMATIONS AND ITS COMPARISON TO THAT OF STRONG FLUCTUATION THEORY

Robert M Manning
Communications Division
National Aeronautics and Space Administration
Cleveland, OH 44135
Robert.M.Manning@nasa.gov

## ABSTRACT

*An expression for the mutual coherence function (MCF) of an electromagnetic beam wave propagating through atmospheric turbulence is derived within the confines of the Rytov approximation. It is shown that both the first and second Rytov approximations are required. The Rytov MCF is then compared to that which issues from the parabolic equation method of strong fluctuation theory. The agreement is found to be quite good in the weak fluctuation case. However, an instability is observed for the special case of beam wave intensities. The source of the instabilities is identified to be the characteristic way beam wave amplitudes are treated within the Rytov method.*

## I. INTRODUCTION

In the early studies in the late 1950's of electromagnetic wave propagation through atmospheric turbulence [1,2], the weak fluctuation theory known as the Rytov approximation made considerable strides in the understanding of the scattering mechanisms inherent in the random permittivity field of the troposphere. Within the structure of the theory, propagation quantities such as the log-amplitude and phase fluctuations took precedence as well as their characterizing statistical parameters, i.e., the corresponding correlation and structure functions of these quantities. Only in the special cases of plane and spherical waves were these quantities related to the prevailing statistical parameters of the electric field of the wave, i.e., the second order statistical moment known as the mutual coherence function (MCF). The case of beam wave propagation was examined [3] but an expression for the MCF was never given. The theory, however, had its limitations and was replaced with the strong-fluctuation theories based on the parabolic equation [2]. In this treatment, the relevant propagation quantities were necessarily the statistical moments of the electric field of the propagating electromagnetic wave; the concepts of log-amplitude and phase fluctuations did not, nor needed not, enter into the structure of strong fluctuation theory. Furthermore, since the MCF was the statistically most significant quantity for applications and experiments, strong fluctuation theory considered not only this quantity for the special cases of plane and spherical wave propagation but also for the more general beam wave case. It was found, without explanation, that the results for the MCF of the plane and spherical wave cases as calculated from the parabolic equation method were identical to those from the Rytov approximation. Due to the success of the parabolic equation method, Rytov theory fell by the wayside and the calculation of the MCF for the beam wave case was never carried-out within weak-fluctuation theory and a comparison was never made to that of strong-fluctuation theory.

It is the purpose of the present study is to essentially finish bridging the gulf between the Rytov approximation and the parabolic equation method by considering the MCF of the beam wave case within the Rytov approximation and comparing it to the results of strong fluctuation theory. By definition, the weak fluctuation theoretical results should directly follow from those of the strong fluctuation results in the limit of weak fluctuations. This will be shown for the beam wave scenario. In the interim, it will also be shown that the traditional form of Rytov theory used in the classical studies is incomplete [4]; the subtleties between the order of the Rytov approximation and the order of magnitude of the permittivity fluctuations was overlooked for a complete and unified application of the theory. Finally, a surprising limitation of the beam wave modeling within Rytov theory is identified and studied.

## II. RECAPITULATION OF THE RYTOV APPROXIMATIONS

Using the field decomposition for the electric field of a wave propagating predominantly along the $x$-axis of a coordinate system situated in a random medium, viz.,

$$E(\vec{r}) = U(\vec{r})\exp(ikx) \qquad (1)$$

in the stochastic Helmholtz equation, characterized by the stochastic permittivity field $\tilde{\varepsilon}(\vec{r})$,

$$\nabla^2 E(\vec{r}) + k^2 E(\vec{r}) + k^2 \tilde{\varepsilon}(\vec{r}) E(\vec{r}) = 0 \qquad (2)$$

one has for the complex amplitude

$$\nabla_\rho^2 U(\vec{r}) + \frac{\partial^2 U(\vec{r})}{\partial x^2} + 2ik\frac{\partial U(\vec{r})}{\partial x} + k^2 \tilde{\varepsilon}(\vec{r})U(\vec{r}) = 0. \qquad (3)$$

In the event that the wavelength $\lambda$ of the propagating wave and the size of the smallest inhomogeneity $l_0$ characterizing the stochastic permittivity field is such that $\lambda \ll l_0$, one has that

$$\frac{\partial^2 U(\vec{r})}{\partial x^2} \ll \left| 2ik\frac{\partial U(\vec{r})}{\partial x} \right| \qquad (4)$$

allowing Eq.(3) to be approximately written

RMManning6/6/08

$$\nabla_\rho^2 U(\bar{r}) + 2ik\frac{\partial U(\bar{r})}{\partial x} + k^2 \tilde{\varepsilon}(\bar{r})U(\bar{r}) = 0 \tag{5}$$

which is a parabolic differential equation of the diffusion type.

The most straightforward way of dealing with this scenario is to employ Tatarskii's method of solution employing the Rytov transformation [2]. The idea is to decouple the stochastic factor $\tilde{\varepsilon}(\bar{r})$ from the resulting stochastic field $U(\bar{r})$ in Eq.(5). To this end, one can employ the transformation (the Rytov transformation)

$$U(\bar{r}) = \exp(\psi(\bar{r})) \tag{6}$$

in Eq.(5) and obtain

$$2ik\frac{\partial \psi(\bar{r})}{\partial x} + \nabla_\rho^2 \psi(\bar{r}) + (\bar{\nabla}_\rho \psi(\bar{r}))^2 + k^2\tilde{\varepsilon}(\bar{r}) = 0 \tag{7}$$

which now a non-parametric relation but also a nonlinear one. Solving this equation via a perturbation expansion in the quantity $v = \sqrt{\langle \tilde{\varepsilon}^2 \rangle}$, one has

$$U(\bar{r}) = U_0(\bar{r})\exp(\psi_1(\bar{r}) + \psi_2(\bar{r}) + \cdots) \tag{8}$$

in which $U_0(\bar{r})$ is the initial field distribution and

$$\psi_1(\bar{r}) = -k^2 \int_{-\infty}^{\infty} G_p(\bar{r},\bar{r}')\tilde{\varepsilon}(\bar{r}')\left(\frac{U_0(\bar{r}')}{U_0(\bar{r})}\right)d^3r' \tag{9}$$

is the first Rytov approximation and

$$\psi_2(\bar{r}) = -\int_{-\infty}^{\infty} G_p(\bar{r},\bar{r}')[\bar{\nabla}_\rho \psi_1(\bar{r}')]^2 \left(\frac{U_0(\bar{r}')}{U_0(\bar{r})}\right)d^3r' \tag{10}$$

is the second Rytov approximation. Here, the parabolic equation Green function is

$$G_p(\bar{r},\bar{r}') = -\left(\frac{1}{4\pi}\right)\frac{\exp(ik(\bar{\rho}-\bar{\rho}')^2/(2(x-x')))}{x-x'} \tag{11}$$

Writing the complex amplitude $U(\bar{r})$ in terms of an amplitude $A(\bar{r})$ and a phase $S(\bar{r})$, i.e.,

$$U_0(\bar{r}) = A_0(\bar{r})\exp(iS_0(\bar{r})), \qquad U(\bar{r}) = A(\bar{r})\exp(iS(\bar{r})) \tag{12}$$

and using these representations in Eq.(8) gives

$$\psi_1(\bar{r}) + \psi_2(\bar{r}) + \cdots = \ln\left[\frac{A(\bar{r})}{A_0(\bar{r})}\right] - i(S_0(\bar{r}) - S(\bar{r}))$$

$$= \chi(\bar{r}) + iS_1(\bar{r}) \tag{13}$$

in which $\chi(\bar{r}) \equiv \ln\left[A(\bar{r})/A_0(\bar{r})\right]$ is the log-amplitude and $S_1(\bar{r}) \equiv S(\bar{r}) - S_0(\bar{r})$ is the phase fluctuation. Hence, letting $\Psi(\bar{r}) \equiv \psi_1(\bar{r}) + \psi_2(\bar{r}) + \cdots$, one can write for these propagation parameters

$$\chi(\bar{r}) = \text{Re}\{\psi_1(\bar{r}) + \psi_2(\bar{r}) + \cdots\} = \frac{1}{2}\left(\Psi(\bar{r}) + \Psi^*(\bar{r})\right) \tag{14}$$

and

$$S_1(\bar{r}) = \text{Im}\{\psi_1(\bar{r}) + \psi_2(\bar{r}) + \cdots\} = \frac{1}{2i}\left(\Psi(\bar{r}) - \Psi^*(\bar{r})\right) \tag{15}$$

These random quantities can only be described by statistical functions such as averages, correlation functions, etc. For example, the spatial log-amplitude correlation function is given by

$$\begin{aligned}B_\chi(\bar{r}_1,\bar{r}_2) &\equiv \frac{1}{4}\left\langle\left(\Psi(\bar{r}_1) + \Psi^*(\bar{r}_1)\right)\left(\Psi(\bar{r}_2) + \Psi^*(\bar{r}_2)\right)\right\rangle \\ &= \frac{1}{4}\left\langle\Psi(\bar{r}_1)\Psi(\bar{r}_2) + \Psi(\bar{r}_1)\Psi^*(\bar{r}_2) + \Psi^*(\bar{r}_1)\Psi(\bar{r}_2) + \Psi^*(\bar{r}_1)\Psi^*(\bar{r}_2)\right\rangle \\ &= \frac{1}{2}\text{Re}\left\{\left\langle\Psi(\bar{r}_1)\Psi(\bar{r}_2)\right\rangle + \left\langle\Psi(\bar{r}_1)\Psi^*(\bar{r}_2)\right\rangle\right\}\end{aligned} \tag{16}$$

Similarly, for the spatial phase correlation and log-amplitude/phase cross correlation

$$B_S(\bar{r}_1,\bar{r}_2) = \frac{1}{2}\text{Re}\left\{\left\langle\Psi(\bar{r}_1)\Psi(\bar{r}_2)\right\rangle - \left\langle\Psi(\bar{r}_1)\Psi^*(\bar{r}_2)\right\rangle\right\} \tag{17}$$

and

$$B_{\chi S}(\bar{r}_1,\bar{r}_2) = \frac{1}{2}\text{Im}\left\{\left\langle\Psi(\bar{r}_1)\Psi(\bar{r}_2)\right\rangle - \left\langle\Psi(\bar{r}_1)\Psi^*(\bar{r}_2)\right\rangle\right\} \tag{18}$$

However, at this point, it is necessary to consider the relationship between the order in $|\tilde{\varepsilon}|$ of the statistics of $\psi_i(\bar{r})$ and the order of the Rytov approximation.

## III. CONNECTION BETWEEN THE FIRST AND SECOND ORDER RYTOV APPROXIMATIONS - THE NECESSITY OF USING BOTH

It is assumed, without loss of generality, that the average or statistical mean $\langle \tilde{\varepsilon}(\vec{r}) \rangle = 0$. Thus, by Eq.(9), $\langle \psi_1(\vec{r}) \rangle = 0$. However, by Eq.(10), $\psi_2(\vec{r}) \sim (\tilde{\varepsilon}(\vec{r}))^2$ and its mean is non-zero. That is, one must consider the second order Rytov approximation, which is of second order in the random permittivity, for a non-zero mean field. Hence, one has from Eqs.(14) and (15), to within the second order Rytov approximation (i.e., of second order in the permittivity fluctuations)

$$\langle \chi(\vec{r}) \rangle = \text{Re}\{\langle \psi_2(\vec{r}) \rangle\}, \qquad \langle S_1(\vec{r}) \rangle = \text{Im}\{\langle \psi_2(\vec{r}) \rangle\} \qquad (19)$$

Now the correlation functions formed using $\psi_1(\vec{r})$ as required by Eqs.(16)-(18), e.g., $\langle \psi_1(\vec{r}_1)\psi_1(\vec{r}_2) \rangle$ are also of second order in the permittivity fluctuations by Eq.(9). Hence, again, to within second order in the permittivity fluctuations, only the first order Rytov approximation need be used for these quantities. Thus, for a solution to within second order in $|\tilde{\varepsilon}|$, by Eqs.(14) and (15), one need only keep the $\psi_1(\vec{r})$ in the evaluation of the spatial correlation functions. Hence, Eqs.(14) and (15) reduce to

$$\chi(\vec{r}) = \text{Re}\{\psi_1(\vec{r})\} = \frac{1}{2}(\psi_1(\vec{r}) + \psi_1^*(\vec{r})), \quad S_1(\vec{r}) = \text{Im}\{\psi_1(\vec{r})\} = \frac{1}{2i}(\psi_1(\vec{r}) - \psi_1^*(\vec{r})) \qquad (20)$$

where

$$B_\chi(\vec{r}_1, \vec{r}_2) = \frac{1}{2}\text{Re}\{\langle \psi_1(\vec{r}_1)\psi_1^*(\vec{r}_2) \rangle + \langle \psi_1(\vec{r}_1)\psi_1(\vec{r}_2) \rangle\} \qquad (21)$$

and

$$B_S(\vec{r}_1, \vec{r}_2) = \frac{1}{2}\text{Re}\{\langle \psi_1(\vec{r}_1)\psi_1^*(\vec{r}_2) \rangle - \langle \psi_1(\vec{r}_1)\psi_1(\vec{r}_2) \rangle\} \qquad (22)$$

Similarly, for the cross correlation of log-amplitude and phase fluctuations

$$B_{\chi S}(\vec{r}_1, \vec{r}_2) = \frac{1}{2}\text{Im}\{\langle \psi_1(\vec{r}_1)\psi_1(\vec{r}_2) \rangle - \langle \psi_1(\vec{r}_1)\psi_1^*(\vec{r}_2) \rangle\} \qquad (23)$$

Therefore, only the product averages $\langle \psi_1(\vec{r}_1)\psi_1(\vec{r}_2) \rangle$ and $\langle \psi_1(\vec{r}_1)\psi_1^*(\vec{r}_2) \rangle$ need to be evaluated for these correlation functions. Of course, from these correlation functions come the associated variances $\sigma_\chi^2(\vec{r}) \equiv B_\chi(\vec{r}, \vec{r})$, $\sigma_S^2(\vec{r}) \equiv B_S(\vec{r}, \vec{r})$, and $\sigma_{\chi S}(\vec{r}) \equiv B_{\chi S}(\vec{r}, \vec{r})$. Finally, the structure functions $D_\chi(\vec{r}_1, \vec{r}_2) \equiv \langle (\chi(\vec{r}_1) - \chi(\vec{r}_2))^2 \rangle$, etc., can be calculated using Eq.(20). Both the mean values given by Eq.(19) and the correlation functions of Eqs.(20)-(23) are second order quantities in the permittivity fluctuations $|\tilde{\varepsilon}|$.

What is important in the foregoing is the salient fact that the first moment given by the second order Rytov approximation, being $\sim (\tilde{\varepsilon}(\bar{r}))^2$, *is just as statistically significant* as those correlation and structure functions found using the first order approximation [4], [5]. Any application requiring the correlation and structure functions *is incomplete* without the consideration of the first order moments $\langle \chi \rangle$ and $\langle S_1 \rangle$. This circumstance will be demonstrated by the application of Rytov theory for the calculation of the MCF of the propagating wave field.

Complete expressions for the log-amplitude and phase fluctuation statistics of a beam wave are given in Appendix A.

## IV. THE MUTUAL COHERENCE FUNCTION WITHIN THE RYTOV APPROXIMATION

In the original applications of Rytov theory to stochastic wave propagation, correlation and structure functions were calculated and compared with experimental results. The MCF, i.e., the second order moment of the associated electric field was relegated to strong fluctuation theory and the parabolic equation method. However, with the proper accounting of the second Rytov approximation, an MCF expression can be derived based on Rytov theory.

Using Eq.(1), the MCF for a beam wave is defined by

$$\Gamma_2(L,\bar{\rho}_1,\bar{\rho}_2) \equiv \langle E(L,\bar{\rho}_1)E^*(L,\bar{\rho}_2) \rangle = \langle U(L,\bar{\rho}_1)U^*(L,\bar{\rho}_2) \rangle \qquad (24)$$

Substituting Eq.(13) into Eq.(8) and using this intermediate result in Eq.(24) yields

$$\Gamma_2(L,\bar{\rho}_1,\bar{\rho}_2) = U_0(L,\bar{\rho}_1)U_0^*(L,\bar{\rho}_2) \langle \exp[\chi(L,\bar{\rho}_1) + \chi(L,\bar{\rho}_2) + i(S_1(L,\bar{\rho}_1) - S_1(L,\bar{\rho}_1))] \rangle \qquad (25)$$

At this point, it is advantageous to recognize that the ensemble average as indicated in Eq.(25) can be written in terms of the corresponding characteristic functional

$$\langle \exp[\chi(L,\bar{\rho}_1) + \chi(L,\bar{\rho}_2) + i(S_1(L,\bar{\rho}_1) - S_1(L,\bar{\rho}_1))] \rangle =$$
$$= \langle \exp[iq\{\chi(L,\bar{\rho}_1) + \chi(L,\bar{\rho}_2) + i(S_1(L,\bar{\rho}_1) - S_1(L,\bar{\rho}_1))\}] \rangle \Big|_{q=-i} \qquad (26)$$

At this point, it is convenient for notational purposes to let $\chi(L,\bar{\rho}_1) \equiv \chi(1)$, etc., and to write $\chi(1) = \chi_1(1) + \chi_2(1)$ where $\chi_1(1)$ is the first order Rytov approximation for $\chi(1)$ and $\chi_2(1)$ for the second order approximation of the quantity; similarly for $S(1) = S_1(1) + S_2(1)$. Since, as mentioned above, quantities up to the second order in the fluctuations $|\tilde{\varepsilon}|$ are only being considered, one, can perform a cumulant expansion of the right side of Eq.(26) up to second order in the parameter $q$ and obtain [6],

$$\left\langle \exp\left[iq\{\chi(L,\bar{\rho}_1) + \chi(L,\bar{\rho}_2) + i(S_1(L,\bar{\rho}_1) - S_1(L,\bar{\rho}_1))\}\right]\right\rangle\bigg|_{q=-i} =$$

$$= \exp\left[iqK_1 - \frac{1}{2}q^2K_2 + \cdots\right]\bigg|_{q=-i}$$

$$= \exp\left[K_1 + \frac{1}{2}K_2\right] \quad (27)$$

where the first cumulant $K_1$ is given by

$$K_1 = \left\langle \chi_1(1) + \chi_2(1) + \chi_1(2) + \chi_2(2) + i\{S_1(1) + S_2(1)\} - i\{S_1(2) + S_2(2)\}\right\rangle \quad (28)$$

and the second by

$$K_2 = \left\langle \left(\chi_1(1) + \chi_2(1) + \chi_1(2) + \chi_2(2) + i\{S_1(1) + S_2(1)\} - i\{S_1(2) + S_2(2)\}\right)^2\right\rangle -$$

$$- \left\langle \chi_1(1) + \chi_2(1) + \chi_1(2) + \chi_2(2) + i\{S_1(1) + S_2(1)\} - i\{S_1(2) + S_2(2)\}\right\rangle^2 \quad (29)$$

From the discussion in Section III, Eq.(28) reduces to the simple result

$$K_1 = 2\langle \chi_2 \rangle \quad (30)$$

upon remembering that $\langle \chi_2 \rangle$ and $\langle S_2 \rangle$ are independent of transverse position (as indicated by Eq.(A13)). Expanding the square terms in Eq.(29) and ignoring terms of order higher than $|\tilde{\varepsilon}|^2$ gives

$$K_2 = \left\langle (\chi(1) + \chi(2))^2\right\rangle + 2i\left\langle (\chi(1) + \chi(2))(S(1) - S(2))\right\rangle - \left\langle (S(1) - S(2))^2\right\rangle \quad (31)$$

Finally, substituting Eqs.(30) and (31) into Eq.(27) allows ensemble average in Eq.(25) to be written

$$\langle \exp[\cdots]\rangle = \exp\left[\frac{1}{2}\left\langle (\chi(1) + \chi(2))^2\right\rangle + i\left\langle (\chi(1) + \chi(2))(S(1) - S(2))\right\rangle - \frac{1}{2}\left\langle (S(1) - S(2))^2\right\rangle + 2\langle \chi \rangle\right]$$

(32)

By definition, the phase structure function is

$$D_S(1,2) \equiv \left\langle (S(1) - S(2))^2\right\rangle \quad (33)$$

But

$$\left\langle (\chi(1)+\chi(2))^2 \right\rangle = \sigma_\chi^2(1) + 2B_\chi(1,2) + \sigma_\chi^2(2) \tag{34}$$

found by expanding the square within the average. However, again by definition the log-amplitude structure function is

$$D_\chi(1,2) \equiv \left\langle (\chi(1)-\chi(2))^2 \right\rangle = \sigma_\chi^2(1) - 2B_\chi(1,2) + \sigma_\chi^2(2) \tag{35}$$

so

$$\left\langle (\chi(1)+\chi(2))^2 \right\rangle = 2\sigma_\chi^2(1) + 2\sigma_\chi^2(2) - D_\chi(1,2) \tag{36}$$

Finally,

$$\left\langle (\chi(1)+\chi(2))(S(1)-S(2)) \right\rangle = \sigma_{\chi S}(1) - B_{\chi S}(1,2) + B_{\chi S}(2,1) - \sigma_{\chi S}(2) \tag{37}$$

where it is noted that, in the general case, according to Eqs.(A9) and (A11), $B_{\chi S}(1,2) \neq B_{\chi S}(2,1)$ and $\sigma_{\chi S}(1) \neq \sigma_{\chi S}(2)$. Using Eqs.(26), and (27) in Eq.(32) gives

$$\left\langle \exp[\cdots] \right\rangle = \exp\left[-\frac{1}{2}D_W(1,2) + \sigma_\chi^2(1) + \sigma_\chi^2(2) + 2\langle\chi\rangle + i\{\sigma_{\chi S}(1) - \sigma_{\chi S}(2) - B_{\chi S}(1,2) + B_{\chi S}(2,1)\}\right] \tag{38}$$

where $D_W(1,2) = D_\chi(1,2) + D_S(1,2)$ is the wave structure function.

    An interesting occurrence of a linear combination of the cross log-amplitude/phase correlations appears as an imaginary term in Eq.(38). Again, for the general beam wave case, this term does not vanish. Using Eqs.(A9) and (A11), one has for this linear combination

$$\Delta_{\chi S} \equiv \sigma_{\chi S}(1) - \sigma_{\chi S}(2) - B_{\chi S}(1,2) + B_{\chi S}(2,1) = -(2\pi)^2 \left(\frac{k^2}{8}\right) \int_0^L \int_0^\infty \mathrm{Im}\{J_0(\kappa P(2,1)) - J_0(\kappa P(1,2))\} \cdot$$

$$\cdot \exp\left[-\frac{\kappa^2}{k}(L-x)\gamma_I\right] \Phi_\varepsilon(\kappa)\kappa d\kappa dx \tag{39}$$

where $P(i,j) \equiv \left|\gamma\bar{\rho}_i - \gamma^*\bar{\rho}_j\right|$. This statistical parameter as well as all the other relevant parameters are evaluated in Appendix B for the Kolmogorov spectrum of turbulent fluctuations. Substituting these results into Eq.(38) gives

$$\langle \exp[\cdots] \rangle = \exp\left[ -4.352 k^2 C_n^2 \int_0^L \left( \frac{L-x}{k} \gamma_I \right)^{5/6} \mathrm{Re}\left\{ {}_1F_1\left( -\frac{5}{6}, 1; -\frac{kP_{12}^2}{4(L-x)\gamma_I} \right) \right\} dx \right.$$

$$+ i(2.176) k^2 C_n^2 \int_0^L \left( \frac{L-x}{k} \gamma_I \right)^{5/6} \mathrm{Im}\left\{ {}_1F_1\left( -\frac{5}{6}, 1; -\frac{kP_{21}^2}{4(L-x)\gamma_I} \right) \right.$$

$$\left. \left. - {}_1F_1\left( -\frac{5}{6}, 1; -\frac{kP_{12}^2}{4(L-x)\gamma_I} \right) \right\} dx \right] \tag{40}$$

Equation (40) can be simplified considerably. To this end, one notes that

$$P^2(1,2) \equiv P_{12}^2 = P_{21}^{2*} \tag{41}$$

In addition to this, the integral representation of the confluent hypergeometric function

$${}_1F_1(a,b;z) = \frac{2^{1-b} \exp(z/2)}{B(a,b-a)} \int_{-1}^{1} (1-t)^{b-a-1}(1+t)^{a-1} \exp(zt/2) dt \tag{42}$$

yields the fact that

$${}_1F_1(a,b;z^*) = {}_1F_1^*(a,b;z) \tag{43}$$

for real $a$ and $b$. Adopting the notation

$${}_1F_1(P_{12}^2) \equiv {}_1F_1\left( -\frac{5}{6}, 1; -\frac{kP_{12}^2}{4\gamma_I(L-x)} \right)$$

one has from the above considerations,

$${}_1F_1(P_{12}^2) = {}_1F_1^*(P_{21}^2) \tag{44}$$

Thus, it is found that (using $\mathrm{Im}\{z\} = (1/2i)[z - z^*]$)

$$\mathrm{Im}\left\{ {}_1F_1(P_{21}^2) - {}_1F_1(P_{12}^2) \right\} = \frac{1}{i}\left[ {}_1F_1^*(P_{12}^2) - {}_1F_1(P_{12}^2) \right] \tag{45}$$

and

$$\mathrm{Re}\left\{ {}_1F_1(P_{12}^2) \right\} = \frac{1}{2}\left[ {}_1F_1(P_{12}^2) + {}_1F_1^*(P_{12}^2) \right] \tag{46}$$

Substituting Eqs.(45) and (46) into Eq.(40) finally yields the simplified result

$$\langle \exp[\cdots] \rangle = \exp\left[ -4.352 k^2 C_n^2 \int_0^L \left( \gamma_I(x) \frac{L-x}{k} \right)^{5/6} {}_1F_1\left( -\frac{5}{6}, 1; -\frac{k P_{21}^2}{4\gamma_I(x)(L-x)} \right) dx \right]$$

(47)

As required by Eq.(25), one now needs to consider the product of the initial fields of a beam wave; using Eq.(A1), one has

$$U_0(L,\bar{\rho}_1) U_0^*(L,\bar{\rho}_2) = \frac{W_0^2}{W^2} \exp\left[ -\frac{k}{2}\left(\frac{W_0^2}{W^2}\right)\left\{ \alpha_1(\rho_1^2 + \rho_2^2) + \right.\right.$$
$$\left.\left. + i\left[\alpha_2 - (\alpha_1^2 + \alpha_2^2)L\right](\rho_1^2 - \rho_2^2) \right\} \right]$$

(48)

where $W^2 = W_0^2\left[(1-\alpha_2 L)^2 + \alpha_1^2 L^2\right]$ is the beam radius at a distance $L$ from the output aperture. Finally, using equations (47) and (48) in Eq.(25) gives for the MCF for a beam wave, to within the second Rytov approximation, propagating through Kolmogorov turbulence

$$\Gamma_2(L,\bar{\rho}_1,\bar{\rho}_2) = \frac{W_0^2}{W^2} \exp\left[ -\frac{k}{2}\left(\frac{W_0^2}{W^2}\right)\left\{ \alpha_1(\rho_1^2 + \rho_2^2) + i\left[\alpha_2 - (\alpha_1^2 + \alpha_2^2)L\right](\rho_1^2 - \rho_2^2) \right\} \right] \cdot$$
$$\cdot \exp\left[ -4.352 k^2 C_n^2 \int_0^L \left( \gamma_I(x) \frac{L-x}{k} \right)^{5/6} {}_1F_1\left( -\frac{5}{6}, 1; -\frac{k P_{21}^2}{4\gamma_I(x)(L-x)} \right) dx \right]$$

(49)

To make contact with other formulations in what is to follow, it is advantageous to place this result in terms of the sum and difference coordinates of $\bar{\rho}_1$ and $\bar{\rho}_2$, viz,

$$\bar{\rho}_c \equiv \frac{\bar{\rho}_1 + \bar{\rho}_2}{2}, \quad \bar{\rho}_d \equiv \bar{\rho}_2 - \bar{\rho}_1$$

(50)

Solving these for $\bar{\rho}_1$ and $\bar{\rho}_2$ and substituting into Eq.(49) and using the definition for $P_{21}^2$ gives for the beam wave MCF to within the second order Rytov approximation

$$\Gamma_2(L,\bar{\rho}_c,\bar{\rho}_d) = \frac{W_0^2}{W^2} \exp\left[ -\frac{k}{2}\left(\frac{W_0^2}{W^2}\right)\left\{ 2\alpha_1\left(\rho_c^2 + \frac{\rho_d^2}{4}\right) - 2i\left[\alpha_2 - (\alpha_1^2 + \alpha_2^2)L\right](\bar{\rho}_c \cdot \bar{\rho}_d) \right\} \right] \cdot$$
$$\cdot \exp\left[ -4.352 k^2 C_n^2 \int_0^L \left( \gamma_I(x) \frac{L-x}{k} \right)^{5/6} {}_1F_1\left( -\frac{5}{6}, 1; -\frac{k|\gamma_R \bar{\rho}_d - 2i\gamma_I \bar{\rho}_c|^2}{4\gamma_I(x)(L-x)} \right) dx \right] \quad (51)$$

## V. RELATIONSHIP OF THE RYTOV MCF WITH THAT OF STRONG FLUCTUATION THEORY

It will now be demonstrated that the result of Eq.(51) agrees with the weak fluctuation approximation of the similar result obtained using the parabolic equation method of strong fluctuation theory. First, however, the plane and spherical wave cases will be considered. In the plane wave case one has $W_0 \to \infty$ and $R_0 \to \infty$. Thus, $\alpha_1 = 0$, $\alpha_2 = 0$, $\gamma_1(\eta) = 1$ and $\gamma_2(\eta) \to 0$. In this limit, the asymptotic expansion of the confluent hypergeometric function yields

$$\lim_{\gamma_2 \to 0} {}_1F_1\left(-\frac{5}{6}, 1; \frac{-k\rho_d^2}{4\gamma_2(\eta)(L-\eta)}\right) = \frac{1}{\Gamma(11/6)}\left(\frac{k\rho_d^2}{4\gamma_2(\eta)(L-\eta)}\right)^{5/6} \tag{52}$$

Using this expression in Eq.(51) gives for a plane wave

$$\Gamma_{2,pw}(\bar{\rho}_c, \bar{\rho}_d, L) = \Gamma_{2,pw}(\bar{\rho}_d, L) = \exp\left[-1.457 k^2 C_n^2 \rho_d^{5/3} L\right] \tag{53}$$

The spherical wave case is defined by $W_0 \to 0$ and $R_0 \to \infty$. Hence, $\alpha_1 \to \infty$, $\alpha_2 = 0$, $\gamma_1(\eta) = \eta/L$ and $\gamma_2(\eta) \to 0$. The same asymptotic form of Eq.(52) applies and Eq.(51) in this case gives

$$\Gamma_{2,sw}(\bar{\rho}_c, \bar{\rho}_d, L) = \left(\frac{kW_0^2}{2L}\right)^2 \exp\left[-\frac{ik}{L}\bar{\rho}_c \cdot \bar{\rho}_d\right] \exp\left[-1.457 k^2 C_n^2 \rho_d^{5/3} \int_0^L (\eta/L)^{5/3} d\eta\right]$$

$$= \left(\frac{kW_0^2}{2L}\right)^2 \exp\left[-\frac{ik}{L}\bar{\rho}_c \cdot \bar{\rho}_d\right] \exp\left[-0.546 k^2 C_n^2 \rho_d^{5/3} L\right] \tag{54}$$

Considering the case in which $\rho_c = 0$ (i.e., the MCF about the propagation axis) and normalizing this result with respect to $\Gamma_{2,sw}(0,0,L)$ gives

$$\Gamma_{2,sw,norm}(0, \bar{\rho}_d, L) = \exp\left[-0.546 k^2 C_n^2 \rho_d^{5/3} L\right] \tag{55}$$

Thus, Eq.(51) gives the same results as does strong fluctuation theory in these limits.
Now for the comparison to the general beam wave case. Consider the general solution for the beam wave MCF as calculated within strong-fluctuation theory, given by Eq.(20-73) of [7], i.e.,

$$\Gamma_2(\bar{\rho}_c, \bar{\rho}_d, L) = \frac{W_0^2}{8\pi} \int_{-\infty}^{\infty} \exp\left[-a\rho_d^2 - b\kappa_d^2 + c\bar{\rho}_d \cdot \bar{\kappa}_d + i\bar{\rho}_c \cdot \bar{\kappa}_d - H(\bar{\rho}_d, \bar{\kappa}_d, L)\right] d^2\kappa_d \tag{56}$$

where

$$a \equiv \frac{1}{2W_0^2}\left(1+\frac{\alpha_2^2}{\alpha_1^2}\right), \quad b \equiv \frac{W^2}{8} = \frac{W_0^2}{8}\left[(\alpha_1 L)^2 + (1-\alpha_2 L)^2\right]$$

(57)

$$c \equiv \frac{1}{2}\left[\alpha_1 L - \frac{\alpha_2}{\alpha_1}(1-\alpha_2 L)\right]$$

$$H(\bar{\rho}_d,\bar{\kappa}_d,L) \equiv 2\pi k^2 \int_0^L \int_{-\infty}^{\infty}\left[1-\exp\left\{-i\bar{\kappa}\cdot\left(\bar{\rho}_d + \frac{\bar{\kappa}_d \eta}{k}\right)\right\}\right]\Phi_n(\bar{\kappa})d^2\kappa\,d\eta \qquad (58)$$

Here, $\Phi_n(\bar{\kappa})$ is the spatial spectrum of refractive index fluctuations. (The definition of $H$ has been slightly modified from that of Ishimaru to facilitate some calculations later in this development.) The problem now reduces to showing that Eq.(56) reduces to Eq.(51) in the limit of weak fluctuations.

**a. Average Intensity ($\bar{\rho}_d = 0$)**

At the outset, it is instructive to consider the slightly simpler case of the average intensity of a beam wave given in strong fluctuation theory by

$$\langle I(\bar{\rho}_c,L)\rangle = \Gamma_2(\bar{\rho}_c,0,L) = \frac{W_0^2}{8\pi}\int_{-\infty}^{\infty}\exp\left[-b\kappa_d^2 + i\bar{\rho}_c\cdot\bar{\kappa}_d - H(0,\bar{\kappa}_d,L)\right]d^2\kappa_d \qquad (59)$$

From Eq.(58), one has for an isotropic turbulent spectral density

$$H(\bar{\kappa}_d,L) = 2\pi k^2 \int_0^L \int_{-\infty}^{\infty}\left[1-\exp\left\{-i\bar{\kappa}\cdot\left(\frac{\bar{\kappa}_d \eta}{k}\right)\right\}\right]\Phi_n(\bar{\kappa})d^2\kappa\,d\eta$$

$$= (2\pi)^2 k^2 \int_0^L \int_0^{\infty}\left[1-J_0\left(\kappa\frac{\kappa_d \eta}{k}\right)\right]\Phi_n(\kappa)\kappa\,d\kappa\,d\eta \qquad (60)$$

Using the Kolmogorov spectral density for refractive index fluctuations

$$\Phi_n(\kappa) = 0.033 C_n^2 \kappa^{-11/3} \qquad (61)$$

in Eq.(60) and evaluating the resulting the $\kappa$-integral using analytic continuation yields

$$H(\kappa_d,L) = 0.547 k^{1/3} C_n^2 L^{8/3} \kappa_d^{5/3} \qquad (62)$$

Since, in the isotropic case, $H(\bar{\kappa}_d,L) = H(\kappa_d,L)$, Eq.(59) becomes in plane polar coordinates

$$\langle I(\bar{\rho}_c,L)\rangle = \frac{W_0^2}{8\pi} \int_0^\infty \int_0^{2\pi} \exp\left[-b\kappa_d^2 + i\rho_c\kappa_d\cos\theta - H(0,\kappa_d,L)\right] d\theta\, \kappa_d\, d\kappa_d$$

$$= \frac{W_0^2}{4} \int_0^\infty \exp\left[-b\kappa_d^2 - H(\kappa_d,L)\right] J_0(\kappa_d\rho_c)\kappa_d\, d\kappa_d \tag{63}$$

The use of Eq.(62) in Eq.(63) results in an integral that cannot be analytically evaluated. However, in the case of weak fluctuations, i.e., $H(\kappa_d,L) \ll 1$, the exponential in Eq.(63) can be expanded to yield

$$\langle I(\bar{\rho}_c,L)\rangle = \frac{W_0^2}{4} \int_0^\infty \exp\left[-b\kappa_d^2\right]\left(1 - H(\kappa_d,L) + \cdots\right) J_0(\kappa_d\rho_c)\kappa_d\, d\kappa_d \tag{64}$$

Using Eq.(62) in this result allows for an analytic evaluation and gives

$$\langle I(\bar{\rho}_c,L)\rangle = \frac{W_0^2}{8}\frac{1}{b}\left\{\exp\left(-\frac{\rho_c^2}{4b}\right) - 0.547\,\Gamma\!\left(\frac{11}{6}\right) k^{1/3} C_n^2 L^{8/3}\left(\frac{1}{b}\right)^{5/6} {}_1F_1\!\left(\frac{11}{6},1;-\frac{\rho_c^2}{4b}\right) + \cdots\right\} \tag{65}$$

Finally, applying the Kummer transformation

$${}_1F_1\!\left(\frac{11}{6},1;-\frac{\rho_c^2}{4b}\right) = \exp\left(-\frac{\rho_c^2}{4b}\right) {}_1F_1\!\left(-\frac{5}{6},1;\frac{\rho_c^2}{4b}\right) \tag{66}$$

and using the definition of $b$ from Eq.(57) in Eq.(65) results in the average intensity in strong-fluctuation theory in the weak-fluctuation limit

$$\langle I(\bar{\rho}_c,L)\rangle = \frac{W_0^2}{W^2}\exp\left(-\frac{2\rho_c^2}{W^2}\right)\left\{1 - 2.91 k^{1/3} C_n^2 L^{8/3}\left(\frac{1}{W}\right)^{5/3} {}_1F_1\!\left(-\frac{5}{6},1;\frac{2\rho_c^2}{W^2}\right) + \cdots\right\} \tag{67}$$

This result should now be compared with that of Eq.(51) from Rytov theory for $\bar{\rho}_d = 0$; in this instance, Eq.(51) becomes

$$\langle I(\bar{\rho}_c,L)\rangle = \frac{W_0^2}{W^2}\exp\left(-\frac{2\rho_c^2}{W^2}\right) \cdot$$

$$\cdot\exp\left[-4.352 k^2 C_n^2 \int_0^L \left(\gamma_2(\eta)\frac{L-\eta}{k}\right)^{5/6} {}_1F_1\!\left(-\frac{5}{6},1;\frac{k\gamma_2(\eta)\rho_c^2}{(L-\eta)}\right) d\eta\right] \tag{68}$$

Using the definitions of $\gamma_2(\eta)$ and $\alpha_1$, the argument of the confluent hypergeometric function gives

$$\frac{k\gamma_2(\eta)\rho_c^2}{(L-\eta)} = \frac{2\rho_c^2}{W_0^2[\alpha_1^2 L^2 + (1-\alpha_2 L)^2]} = \frac{2\rho_c^2}{W^2} \tag{69}$$

where the last result comes from using a relationship in Eq.(57). Thus, this function becomes independent of $\eta$. Equation (68) is then reduced to

$$\langle I(\bar{\rho}_c, L)\rangle = \frac{W_0^2}{W^2}\exp\left(-\frac{2\rho_c^2}{W^2}\right)\cdot$$

$$\cdot\exp\left[-4.352 k^2 C_{n1}^2 F_1\left(-\frac{5}{6}, 1; \frac{2\rho_c^2}{W^2}\right)\int_0^L \left(\gamma_2(\eta)\frac{L-\eta}{k}\right)^{5/6} d\eta\right] \tag{70}$$

within which the $\eta$-integration can now be performed upon using, once again, the definitions of $\gamma_2(\eta)$ and $\alpha_1$ as well as Eq.(57). This gives

$$\int_0^L \left(\gamma_2(\eta)\frac{L-\eta}{k}\right)^{5/6} d\eta = 2^{5/6}\left(\frac{3}{8}\right) k^{-5/3} L^{8/3}\left(\frac{1}{W}\right)^{5/3} \tag{71}$$

Substituting this into Eq.(70) yields the average intensity in the weak-fluctuation limit of Rytov theory

$$\langle I(\bar{\rho}_c, L)\rangle = \frac{W_0^2}{W^2}\exp\left(-\frac{2\rho_c^2}{W^2}\right)\exp\left[-2.91 k^{1/3} C_n^2 L^{8/3}\left(\frac{1}{W}\right)^{5/3} {}_1F_1\left(-\frac{5}{6}, 1; \frac{2\rho_c^2}{W^2}\right)\right] \tag{72}$$

the series expansion of which agrees with the result of Eq.(67) (this fact was first noted in [4]). Hence, the average intensity of a beam wave as calculated from strong fluctuation theory agrees with that of Rytov theory in the limit of weak fluctuations.

    A digression must now be made with regard to the limits of applicability of the first Rytov approximation before graphical depictions of the comparison of general results is given. As derived in [4], the requirement that the second Rytov approximation can be neglected for correlation functions of log-amplitude and phase fluctuations is given by the condition $|\langle\chi\rangle| \ll 1$; for a general beam wave this is

$$|\langle\chi\rangle| = 1.187 C_n^2 k^{7/6} L^{11/6} \text{Re}\left\{(i)^{5/6} {}_2F_1\left(-\frac{5}{6}, \frac{11}{6}; \frac{17}{6}; \frac{i\alpha L}{1+i\alpha L}\right)\right\} \ll 1 \tag{73}$$

Hence, the limits of applicability are seen to be functions of the parameters of the particular beam wave being considered. Take for example, the propagation conditions given by $L = 2500$ m, $C_n^2 = 5\times 10^{-15}$ m$^{2/3}$, and $\lambda = 0.63\times 10^{-6}$ m. In the case of a collimated beam wave ($R_0 \to \infty$), the behavior of Eq.(73) as a function of $W_0$ is displayed in Figure 1. The plot begins at the spherical wave limit and asymptotically reaches the plane wave limit. The function $|\langle \chi \rangle|$ peaks at the value of $W_0 \sim \sqrt{L\lambda} = .04$ m, i.e., that size of the first Fresnel zone. As can be seen, the condition of Eq.(73) qualitatively breaks down around this value.

Graphical depictions and comparisons of the general results of Eq.(51), with $\rho_d = 0$, and Eq.(63) are shown in Figures 2-5. Here, the normalized intensity of the collimated beam wave cases are shown for initial waist radii $W_0$ of 0.0005 m, 0.005 m, 0.05 m, and 0.5 m; the first value is essentially for a spherical wave case and the last value effectively represents the plane wave case. In all cases, the following propagation parameters prevail: $L = 2500$ m, $\lambda = 0.63\ \mu$, and $C_n^2 = 5 \times 10^{-15}$ m$^{-2/3}$. As can be seen from Fig.3, the results of Rytov theory begin to diverge at $\rho_c \approx 0.2$ m. This phenomenon is at its extreme in Figure 4. The results once again coalesce in Figure 5.

Since the fundamental foundation of Rytov theory and its various approximations are based on the parabolic equation, Eq.(5), such intensity divergences cannot arise from this fact since they are not observed within strong fluctuation theory which is also based on the parabolic equation. As will be shown in part b. of this section, stable results are obtained for the MCF across all ranges of beam wave waist radii. In addition, since palatable results for the intensity are obtained for the limiting cases of plane and spherical waves, it is thus doubtful that such divergence phenomena are occurring due to the neglect or mistreatment of some propagation related mechanism or use of the paraxial approximation. Therefore, the beam structure becomes suspect. But again, such is not the case in strong fluctuation theory so whatever the source of the intensity divergences are, it is only peculiar to the way the beam wave structure is treated within Rytov theory. The study and identification of this intensity divergence behavior will form the subject of Section VI.

One can now proceed to examine the complete expression for the MCF as given by Eq.(56) and compare it to that of the Rytov theory, Eq. (51).

**b. MCF**

From what has been gleaned from the examination of the intensity, one should consider the weak fluctuation limit of Eq.(56) from the outset, i.e.,

$$\Gamma_2(\bar{\rho}_c, \bar{\rho}_d, L) = \frac{W_0^2}{8\pi} \int_{-\infty}^{\infty} \exp[-a\rho_d^2 - b\kappa_d^2 + c\bar{\rho}_d \cdot \bar{\kappa}_d + i\bar{\rho}_c \cdot \bar{\kappa}_d] \cdot$$

$$\cdot \{1 - H(\bar{\rho}_d, \kappa_d, L) + \cdots\} d^2\kappa_d$$

$$= \Gamma_2^{(1)}(\bar{\rho}_c, \bar{\rho}_d, L) - \Gamma_2^{(2)}(\bar{\rho}_c, \bar{\rho}_d, L) + \cdots \quad (74)$$

where

$$\Gamma_2^{(1)}(\bar{\rho}_c, \bar{\rho}_d, L) \equiv \frac{W_0^2}{8\pi} \int_{-\infty}^{\infty} \exp[-a\rho_d^2 - b\kappa_d^2 + c\bar{\rho}_d \cdot \bar{\kappa}_d + i\bar{\rho}_c \cdot \bar{\kappa}_d] d^2\kappa_d \quad (75)$$

and

$$\Gamma_2^{(2)}(\bar{\rho}_c,\bar{\rho}_d,L) \equiv \frac{W_0^2}{8\pi}\int_{-\infty}^{\infty}\exp\left[-a\rho_d^2 - b\kappa_d^2 + c\bar{\rho}_d\cdot\bar{\kappa}_d + i\bar{\rho}_c\cdot\bar{\kappa}_d\right]H(\bar{\rho}_d,\bar{\kappa}_d,L)d^2\kappa_d \quad (76)$$

The evaluation of Eq.(75) is straightforward:

$$\Gamma_2^{(1)}(\bar{\rho}_c,\bar{\rho}_d,L) = \frac{W_0^2}{8\pi}\exp(-a\rho_d^2)\int_{-\infty}^{\infty}\exp\left[i(\bar{\rho}_c - ic\bar{\rho}_d)\cdot\bar{\kappa}_d\right]\exp(-b\kappa_d^2)d^2\kappa_d$$

$$= \frac{W_0^2}{4}\exp(-a\rho_d^2)\int_0^{\infty}J_0(|\bar{\rho}_c - ic\bar{\rho}_d|\kappa_d)\exp(-b\kappa_d^2)\kappa_d d\kappa_d$$

$$= \left(\frac{W_0^2}{4}\right)\left(\frac{1}{2b}\right)\exp\left[-\frac{|\bar{\rho}_c - ic\bar{\rho}_d|^2 + 4ab\rho_d^2}{4b}\right] \quad (77)$$

Equation (76) becomes, with the use of Eq.(58),

$$\Gamma_2^{(2)}(\bar{\rho}_c,\bar{\rho}_d,L) \equiv \frac{W_0^2}{8\pi}(2\pi k^2)\int_{-\infty}^{\infty}\int_0^L\int_{-\infty}^{\infty}\exp\left[-a\rho_d^2 - b\kappa_d^2 + c\bar{\rho}_d\cdot\bar{\kappa}_d + i\bar{\rho}_c\cdot\bar{\kappa}_d\right]\cdot$$

$$\cdot\left[1 - \exp\left\{-i\bar{\kappa}\cdot\left(\bar{\rho}_d + \frac{\bar{\kappa}_d\eta}{k}\right)\right\}\right]\Phi_n(\bar{\kappa})d^2\kappa d\eta d^2\kappa_d \quad (78)$$

Here, the $\kappa_d$-integration will be performed first followed by the $\kappa$-integration. To this end, one has

$$\Gamma_2^2(\bar{\rho}_c,\bar{\rho}_d,L) = \frac{W_0^2}{4}k^2\int_{-\infty}^{\infty}\int_0^L\{T_1 + T_2\}\Phi_n(\bar{\kappa})d\eta d^2\kappa \quad (79)$$

where

$$T_1 \equiv \int_{-\infty}^{\infty}\exp\left[-a\rho_d^2 - b\kappa_d^2 + i(\bar{\rho}_c - ic\bar{\rho}_d)\cdot\bar{\kappa}_d\right]d^2\kappa_d \quad (80)$$

and

$$T_2 \equiv \int_{-\infty}^{\infty}\exp\left[-a\rho_d^2 - b\kappa_d^2 + i(\bar{\rho}_c - ic\bar{\rho}_d)\cdot\bar{\kappa}_d\right]\exp\left[-i\bar{\kappa}\cdot\bar{\rho}_d - i\frac{\bar{\kappa}\eta}{k}\bar{\kappa}_d\right]d^2\kappa_d \quad (81)$$

The evaluation of Eq.(80) follows that of Eq.(77):

$$T_1 = 2\pi\left(\frac{1}{2b}\right)\exp\left[-\frac{|\bar{\rho}_c - ic\bar{\rho}_d|^2 + 4ab\rho_d^2}{4b}\right] \tag{82}$$

Similarly, Eq.(81) becomes

$$T_2 = 2\pi\left(\frac{1}{2b}\right)\exp\left(-a\rho_d^2 - i\vec{\kappa}\cdot\bar{\rho}_d\right)\exp\left[-\frac{\left|\bar{\rho}_c - ic\bar{\rho}_d - \frac{\vec{\kappa}\eta}{k}\right|^2}{4b}\right]$$

$$= 2\pi\left(\frac{1}{2b}\right)\exp\left[-\frac{|\bar{\rho}_c - ic\bar{\rho}_d|^2 + 4ab\rho_d^2}{4b}\right]\exp\left[-i\vec{S}\cdot\vec{\kappa} - \frac{\eta^2}{4bk^2}\kappa^2\right] \tag{83}$$

where

$$\vec{S} \equiv \frac{-2i\frac{\eta}{k}(\bar{\rho}_c - ic\bar{\rho}_d) + 4b\bar{\rho}_d}{4b} \tag{84}$$

Substituting Eqs.(82) and (83) into Eq.(79) gives

$$\Gamma_2^2(\bar{\rho}_c, \bar{\rho}_d, L) = \frac{W_0^2}{4}\left(\frac{1}{2b}\right)2\pi k^2 \exp\left[-\frac{|\bar{\rho}_c - ic\bar{\rho}_d|^2 + 4ab\rho_d^2}{4b}\right]\cdot$$

$$\cdot\int_0^L \int_{-\infty}^{\infty}\left[1 - \exp\left(-i\vec{S}\cdot\vec{\kappa} - \frac{\eta^2}{4bk^2}\kappa^2\right)\right]\Phi_n(\vec{\kappa})d^2\kappa\, d\eta \tag{85}$$

Using the isotropic Kolmogorov spectral density given by Eq.(61) yields for the $\kappa$-integration

$$\int_{-\infty}^{\infty}\left[1 - \exp\left(-i\vec{S}\cdot\vec{\kappa} - \frac{\eta^2}{4bk^2}\kappa^2\right)\right]\Phi_n(\vec{\kappa})d^2\kappa = 0.033(2\pi)C_n^2\cdot$$

$$\cdot\int_0^{\infty}\left[1 - J_0(|\vec{S}|\kappa)\exp\left(-\frac{\eta^2}{4bk^2}\kappa^2\right)\right]\kappa^{-8/3}d\kappa$$

$$= (2\pi)(0.033)C_n^2\left(\frac{1}{2}\right)\left(-\Gamma\left(-\frac{5}{6}\right)\right)\left(\frac{\eta^2}{4bk^2}\right)^{5/6}{}_1F_1\left(-\frac{5}{6}, 1; \frac{|\vec{S}|^2 k^2 b}{\eta^2}\right) \tag{86}$$

where analytic continuation was used to obtain the final result. Using this in Eq.(85) and substituting Eqs.(77) and (85) into Eq.(74) finally yields

$$\Gamma_2(\bar{\rho}_c,\bar{\rho}_d,L) = \frac{W_0^2}{4}\left(\frac{1}{2b}\right)\exp\left[-\frac{|\bar{\rho}_c - ic\bar{\rho}_d|^2 + 4ab\rho_d^2}{4b}\right]\cdot$$
$$\left(1 - 4.351 k^2 C_n^2 \int_0^L \left(\frac{\eta^2}{4bk^2}\right)^{5/6} {}_1F_1\left(-\frac{5}{6},1;\frac{|\bar{S}|^2 k^2 b}{\eta^2}\right)d\eta + \cdots\right) \quad (87)$$

Using the definitions for $a$, $b$, $c$, and $\bar{S}$, and employing the transcription $\eta \to L - \eta$, one obtains, after some rather involved algebra,

$$\Gamma_2(\bar{\rho}_c,\bar{\rho}_d,L) = \left(\frac{1}{(\alpha_1 L)^2 + (1-\alpha_2)^2}\right)\exp\left[-\frac{\alpha_1\left(\rho_c^2 + \frac{\rho_d^2}{4}\right) + i(\alpha_2 - (\alpha_1^2 + \alpha_2^2)L)\bar{\rho}_c \cdot \bar{\rho}_d}{(\alpha_1 L)^2 + (1-\alpha_2)^2}\right]\cdot$$
$$\left(1 - 4.351 k^2 C_n^2 \int_0^L \left(-\gamma_2(\eta)\frac{L-\eta}{k}\right)^{5/6} {}_1F_1\left(-\frac{5}{6},1;\frac{k|\gamma_1\bar{\rho}_d - 2i\gamma_2\bar{\rho}_c|}{4\gamma_2(L-\eta)}\right)d\eta + \cdots\right) \quad (88)$$

The first two factors of this expression describe the overall beam wave structure; what is important to note here, however, is that the expansion given by the third term agrees with the corresponding weak fluctuation expansion of Eq.(51).

Graphs of the normalized MCF's of the entire expression of Eq.(51) with that of Eq.(56) are shown in Figs. 6-8 for a collimated beam wave subtending the spherical to plane wave limits. The largest discrepancy between these results occurs at $W_0 = 0.05$ m; However, the Rytov results do not diverge as they do for the associated intensity of Fig. 4.

### VI. LIMITS OF BEAM WAVE PARAMETERS IMPOSED BY THEIR USE IN THE RYTOV APPROXIMATION

As shown in the last section, the calculation within the first and second Rytov approximations of the intensity of a beam wave propagating through turbulence revealed a divergent instability in the predictions for beam waves of waist radius on the order of the prevailing Fresnel zone. The MCF, however, remained stable in this situation. It thus becomes important to identify the source of this divergence for such beam waves.

As shown in previous treatments, all orders of approximation of Rytov theory possess the combination of initial field distributions

$$R(L,\bar{\rho};x,\bar{\rho}') \equiv \frac{U_0(x,\bar{\rho}')}{U_0(L,\bar{\rho})} \quad (89)$$

where for a Gaussian beam wave, one has Eqs.(A1) and (A2), viz.

$$U_0(x, \bar{\rho}') = \frac{1}{1+i\alpha x} \exp\left[-\left(\frac{k\alpha}{2}\right) \frac{\rho'^2}{1+i\alpha x}\right] \quad (90)$$

with

$$\alpha \equiv \alpha_1 + i\alpha_2, \quad \alpha_1 \equiv \frac{2}{kW_0^2}, \quad \alpha_2 \equiv \frac{1}{R_0} \quad (91)$$

For simplicity, and without loss of generality, consider the collimated beam wave case in which $R_0 \to \infty$, i.e., $\alpha_2 = 0$. Substituting Eq.(90) into Eq.(89) gives

$$R(L, \bar{\rho}; x, \bar{\rho}') = \frac{1+i\alpha_1 L}{1+i\alpha_1 x} \exp\left[-\frac{k\alpha_1}{2}\left\{\frac{\rho'^2}{1+i\alpha_1 x} - \frac{\rho^2}{1+i\alpha_1 L}\right\}\right]$$

$$= \frac{1+i\alpha_1 L}{1+i\alpha_1 x} \exp\left[-\frac{k}{2}\left\{\frac{\rho'^2}{1/\alpha_1 + ix} - \frac{\rho^2}{1/\alpha_1 + iL}\right\}\right]$$

$$= \frac{1+i\alpha_1 L}{1+i\alpha_1 x} \exp\left[-\frac{k}{2}\left\{\frac{\rho'^2(1/\alpha_1 - ix)}{(1/\alpha_1)^2 + x^2} - \frac{\rho^2(1/\alpha_1 - iL)}{(1/\alpha_1)^2 + L^2}\right\}\right]$$

$$= \frac{1+i\alpha_1 L}{1+i\alpha_1 x} \exp\left[A + \frac{ik}{2} S\right] \quad (92)$$

where

$$A \equiv -\frac{k}{2}\left[\frac{1/\alpha_1}{(1/\alpha_1)^2 + x^2} \rho'^2 - \frac{1/\alpha_1}{(1/\alpha_1)^2 + L^2} \rho^2\right] \quad (93)$$

is the amplitude factor describing the evolution of the wave amplitude and

$$S \equiv \frac{x}{(1/\alpha_1)^2 + x^2} \rho'^2 - \frac{L}{(1/\alpha_1)^2 + L^2} \rho^2 \quad (94)$$

is the phase factor.

An important condition for the convergence of any integral expression in which Eq.(92) finds itself is $A < 0$. Thus, consider the amplitude term within the exponential of Eq.(92) more closely, i.e.,

$$A \equiv -\frac{k}{2}K(x)\rho'^2 + \frac{k}{2}K(L)\rho^2, \qquad K(z) \equiv \frac{\left(1/\alpha_1\right)}{\left(1/\alpha_1\right)^2 + z^2} \tag{95}$$

Look now at the first term of on the right side of Eq.(95) and consider the two cases $(1/\alpha_1)^2 \gg x^2$ and $(1/\alpha_1)^2 \ll x^2$, i.e., using the definition of $\alpha_1$, $W_0 > \sqrt{\lambda x}$ and $W_0 < \sqrt{\lambda x}$.

(i) $(1/\alpha_1)^2 \gg x^2$

Take the worst case in which $x = L$. Thus, $K(x) \sim K(L) = \alpha_1$ and one has for the first term on Eq.(95)

$$\frac{k}{2}K(x)\rho'^2 \sim \frac{k}{2}K(L)\rho'^2 = \frac{k}{2}\alpha_1\rho'^2 = \frac{\rho'^2}{W_0^2} \tag{96}$$

For any integral over $\rho'$ in which the expression of Eq.(92) occurs, most contributions over the integration range come from $\rho'^2/W_0^2 \leq 1$. Taking the equality to hold in the worst case and using this condition in Eq.(96) yields

$$\frac{k}{2}K(x)\rho'^2 \sim \frac{k}{2}K(L)\rho'^2 \sim 1 \tag{97}$$

Substituting this intermediate result into Eq.(95) gives for the condition of convergence of any integral involving Eq.(92), i.e., $A < 0$,

$$A = -1 + \frac{k}{2}K(L)\rho^2 < 0 \tag{98}$$

(ii) $(1/\alpha_1)^2 \ll x^2$

Once again, taking $x = L$, one has $K(x) = (1/\alpha_1)/L^2$. Hence, for the first term on the right of Eq.(95)

$$\frac{k}{2}K(x)\rho'^2 \sim \frac{k}{2}K(L)\rho'^2 = \frac{k}{2}\frac{1/\alpha_1}{L^2}\rho'^2 = \frac{k^2 W_0^2}{4L^2}\rho'^2 \sim \left(\frac{W_0^2}{\lambda L}\right)\left(\frac{\rho'^2}{\lambda L}\right) \tag{99}$$

But by assumption, $W_0^2 \ll \lambda L$. However, in most propagation scenarios, $\rho' > \sqrt{\lambda L}$ in the significant range of integration over $\rho'$; there is thus an indeterminate size for the last term of Eq.(99). At this point, without further analysis, it will be taken to be $\sim 1$. Therefore, once again the condition

$$\frac{k}{2}K(x)\rho'^2 \sim \frac{k}{2}K(L)\rho'^2 \sim 1 \tag{100}$$

obtains and Eq.(95) gives for a convergent integral condition

$$A = -1 + \frac{k}{2}K(L)\rho^2 < 0 \tag{101}$$

which is identical to the result in Eq.(98) of Case (i).

Using the definitions given above, Eqs.(98) and (101) reduce to

$$\frac{\pi\alpha_1}{1+\alpha_1^2 L^2} < 1 \tag{102}$$

which can be further reduced to

$$\frac{\rho^2/W_0^2}{1+\lambda^2 L^2/\pi^2 W_0^4} < 1 \tag{103}$$

This finally can be simplified once more to yield the constraint

$$\rho^2 < W_0^2 + \frac{\lambda^2 L^2}{\pi^2 W_0^2} \tag{104}$$

existing between the beam wave and propagation quantities.

This condition holds for all types of collimated beam waves and can also be a good indicator for convergent and divergent beam waves. It gives the possible range of values of the transverse coordinate for which the intensity of a beam wave remains stable within Rytov theory. For plane ($W_0 \to \infty$) and spherical waves ($W_0 \to 0$), the condition holds trivially. The condition shows itself for nominal values of $W_0$. For example, as shown in Fig. 3, a case in which the intensity divergence was noted to occur is given by $W_0 = 0.005$ m, $\lambda = 0.63$ μm and $L = 2500$ m. This gives through Eq.(104) $\rho < 0.1$ m which is just before the intensity divergence sets in. However, the case for which $W_0 = 0.05$ m shown in Fig. 4, i.e., right at the Fresnel zone length, gives $\rho < 0.05$ m; the intensity divergence occurred well before this value. The discrepancy is be due to the fact that an asymptotic analysis on either side of the quantity $\sqrt{\lambda L}$ was used to derive Eq.(104). One can expect a disagreement in situations in which $W_0 \sim \sqrt{\lambda L}$. A more complete analysis than that given above is needed to cover such cases. Suffice it to say

RMManning6/6/08

that the condition $W_0 \sim \sqrt{\lambda L}$ precludes the use of Rytov theory to describe the behavior of beam wave intensity.

Thus the source of intensity divergences for beam waves within the confines of Rytov theory has been identified and an expression is given for the stable range of intensity predictions. The fact that the MCF remains stable over these transverse ranges is connected with the fact that the difference coordinate over which the MCF is evaluated is symmetric about the beam wave axis; the centroid coordinate is taken to be zero in these cases. Thus, the divergences that arise off axis are more or less canceled by performing the field comparison across the difference coordinates as is done to obtain the MCF.

## VII. CONCLUSION

A closed form solution for the MCF based on the first and second Rytov approximations has been derived and compared with that from strong fluctuation theory. The agreement between these two approaches is quite good in all beam wave cases that satisfy the weak fluctuation case save for the intensity distribution for a collimated beam wave. In the region where the initial beam waist size is on the order of the first Fresnel zone length, the solution in the Rytov case diverges. The source of this divergence is found to be in the way a beam wave is modeled within the Rytov approach; it arises from the ratio of the complex amplitudes that occurs within the theory.

**APPENDIX A**

One has for a beam wave of waist radius $W_0$ and a phase front radius of curvature $R_0$ both measured at the output aperture of the transmitter, an initial field distribution of unit amplitude given by [2,3]

$$U_0(\bar{r}) = U_0(x,\bar{\rho}) = \frac{1}{1+i\alpha x}\exp\left[-\left(\frac{k\alpha}{2}\right)\frac{\rho^2}{1+i\alpha x}\right] \quad (A1)$$

where

$$\alpha \equiv \alpha_1 + i\alpha_2, \quad \alpha_1 \equiv \frac{2}{kW_0^2}, \quad \alpha_2 \equiv \frac{1}{R_0} \quad (A2)$$

using Eqs.(9) and (11) with Eq.(A1), one obtains the correlation expressions

$$\langle \psi_1(L,\bar{\rho}_1)\psi_1(L,\bar{\rho}_2)\rangle = -\frac{k^2}{4}\int_0^L\int_{-\infty}^{\infty}\exp(i\bar{\kappa}\cdot\bar{Q})H^2(L-x,\kappa)F_\varepsilon(\bar{\kappa})d^2\kappa dx \quad (A3)$$

and

$$\langle \psi_1(L,\bar{\rho}_1)\psi_1^*(L,\bar{\rho}_2)\rangle = \frac{k^2}{4}\int_0^L\int_{-\infty}^{\infty}\exp(i\bar{\kappa}\cdot\bar{P})|H(L-x,\kappa)|^2 F_\varepsilon(\bar{\kappa})d^2\kappa dx \quad (A4)$$

where the complex coordinates are $\bar{Q} \equiv \gamma(\bar{\rho}_1 - \bar{\rho}_2)$ and $\bar{P} \equiv \gamma\bar{\rho}_1 - \gamma^*\bar{\rho}_2$ with

$$\gamma = \gamma(L,x) \equiv \frac{1+i\alpha x}{1+i\alpha L} = \gamma_R - i\gamma_I, \qquad \gamma_I > 0$$

$$\gamma_R \equiv \frac{(1-\alpha_2 x)(1-\alpha_2 L) + \alpha_1^2 Lx}{(1-\alpha_2 L)^2 + \alpha_1^2 L^2}, \quad \gamma_I \equiv \frac{\alpha_1(L-x)}{(1-\alpha_2 L)^2 + \alpha_1^2 L^2} \quad (A5)$$

and

$$H(L-x,\kappa) \equiv \exp\left[-\frac{i\kappa^2}{2k}(L-x)\gamma\right]. \quad (A6)$$

Finally, the two dimensional spectral amplitude of the permittivity fluctuations is given by $F_\varepsilon(\bar{\kappa}) = 2\pi\Phi_\varepsilon(\bar{\kappa})$ where $\Phi_\varepsilon(\bar{\kappa})$ is the associated three-dimensional spectral density.
The solution to the second Rytov approximation yields [4]

$$\langle \psi_2(L,\bar{\rho})\rangle = \langle \psi_2(L)\rangle = -i\frac{k}{8}\int_0^L \int_0^x \int_{-\infty}^{\infty} \gamma^2(x,x')H^2(x-x',\kappa)\kappa^2 F_\varepsilon(\bar{\kappa})d^2\kappa dx'dx \quad (A7)$$

a result that is independent of the transverse position.

From Eqs.(21)-(23), one obtains in the case of an isotropic spatial spectrum of fluctuations [3]

$$B_{\substack{\chi \\ S}}(L,\bar{\rho}_1,\bar{\rho}_2) = (2\pi)^2\left(\frac{k^2}{8}\right)\int_0^L \int_0^\infty \mathrm{Re}\left\{J_0(\kappa P)\exp\left[-\frac{\kappa^2}{k}(L-x)\gamma_I\right]\mp\right.$$
$$\left.\mp J_0(\kappa Q)\exp\left[-\frac{i\kappa^2}{k}(L-x)\gamma\right]\right\}\Phi_\varepsilon(\kappa)\kappa d\kappa dx \quad (A8)$$

and

$$B_{\chi S}(L,\bar{\rho}_1,\bar{\rho}_2) = -(2\pi)^2\left(\frac{k^2}{8}\right)\int_0^L \int_0^\infty \mathrm{Im}\left\{J_0(\kappa P)\exp\left[-\frac{\kappa^2}{k}(L-x)\gamma_I\right]+\right.$$
$$\left.+J_0(\kappa Q)\exp\left[-\frac{i\kappa^2}{k}(L-x)\gamma\right]\right\}\Phi_\varepsilon(\kappa)\kappa d\kappa dx \quad (A9)$$

In particular,

$$\sigma_{\substack{\chi \\ S}}^2(L,\bar{\rho}) = (2\pi)^2\left(\frac{k^2}{8}\right)\int_0^L \int_0^\infty \mathrm{Re}\left\{J_0(2i\gamma_I \kappa\rho)\exp\left[-\frac{\kappa^2}{k}(L-x)\gamma_I\right]\mp\right.$$
$$\left.\mp\exp\left[-\frac{i\kappa^2}{k}(L-x)\gamma\right]\right\}\Phi_\varepsilon(\kappa)\kappa d\kappa dx \quad (A10)$$

and

$$\sigma_{\chi S}(L,\bar{\rho}) = -(2\pi)^2\left(\frac{k^2}{8}\right)\int_0^L \int_0^\infty \mathrm{Im}\left\{J_0(2i\gamma_I \rho\kappa)\exp\left[-\frac{\kappa^2}{k}(L-x)\gamma_I\right]+\right.$$
$$\left.+\exp\left[-\frac{i\kappa^2}{k}(L-x)\gamma\right]\right\}\Phi_\varepsilon(\kappa)\kappa d\kappa dx \quad (A11)$$

In addition, one has for the log-amplitude and phase structure functions

$$D_{\substack{\chi \\ S}}(L,\bar{\rho}_1,\bar{\rho}_2) = (2\pi)^2\left(\frac{k^2}{8}\right)\int_0^L \int_0^\infty \mathrm{Re}\left\{\left(J_0(2i\gamma_I\kappa\rho_1)+J_0(2i\gamma_I\kappa\rho_2)-2J_0(\kappa P)\right)\cdot\right.$$
$$\left.\cdot\exp\left[-\frac{\kappa^2}{k}(L-x)\gamma_I\right]\mp 2(1-J_0(\kappa Q))\exp\left[-\frac{i\kappa^2}{k}(L-x)\gamma\right]\right\}\Phi_\varepsilon(\kappa)\kappa d\kappa dx \quad (A12)$$

Finally, for the mean log-amplitude and phase fluctuations [4]

$$\left.\begin{array}{l}\langle\chi(L)\rangle\\ \langle S(L)\rangle\end{array}\right\} = -(2\pi)^2 \left\{\begin{array}{l}\text{Re}\\ \text{Im}\end{array}\right\}\left(\frac{ik}{8}\right)\int_0^L \int_0^x \int_0^\infty \gamma^2(x,x')\exp\left[-\frac{i\kappa^2}{k}(x-x')\gamma(x,x')\right]\kappa^2 \cdot$$
$$\cdot \Phi_\varepsilon(\kappa)\kappa\, d\kappa\, dx'\, dx \qquad \text{(A13)}$$

In these expressions, $\gamma_I = -\text{Im}\{\gamma\} > 0$. Before an evaluation of these various equations is given for the Kolmogorov spectrum of turbulence, an expression for the MCF will be developed.

## APPENDIX B

Using the Kolmogorov spectrum for permittivity fluctuations $\Phi_\varepsilon(\kappa) = 0.033 C_\varepsilon^2 \kappa^{-11/3}$ where, in terms of the refractive index structure constant $C_\varepsilon^2 = 4C_n^2$, and evaluating the requisite functions needed in Eq.(38), one obtains the following results from Eqs.(39), (A10), (A12), and (A13)

$$D_W(1,2) \equiv D_W(L,\bar{\rho}_1,\bar{\rho}_2) = D_\chi(L,\bar{\rho}_1,\bar{\rho}_2) + D_S(L,\bar{\rho}_1,\bar{\rho}_2) =$$
$$= -0.816 C_n^2 k^{7/6} L^{11/6} \left(\frac{\alpha_1 L}{(1-\alpha_2 L)^2 + \alpha_1^2 L^2}\right)^{5/6} \cdot$$
$$\cdot \left({}_1F_1\left(-\frac{5}{6},1;\frac{2\rho_1^2}{W^2}\right) + {}_1F_1\left(-\frac{5}{6},1;\frac{2\rho_2^2}{W^2}\right)\right) + 4.352 k^2 C_n^2 \int_0^L \left(\left(\frac{L-x}{k}\right)\gamma_I\right)^{5/6} \cdot$$
$$\cdot \text{Re}\left\{{}_1F_1\left(-\frac{5}{6},1;-\frac{kP_{12}^2}{4(L-x)\gamma_I}\right)\right\} dx \qquad \text{(B1)}$$

$$\sigma_\chi^2(L,\rho) = 2.176 k^{7/6} L^{11/6} C_n^2 \left[\text{Re}\left\{\left(\frac{6}{11}\right)i^{5/6}{}_2F_1\left(-\frac{5}{6},\frac{11}{6},\frac{17}{6};\frac{i\alpha L}{1+i\alpha L}\right)\right\} -$$
$$-\left(\frac{3}{8}\right)\left(\frac{\alpha_1 L}{(1-\alpha_2 L)^2 + \alpha_1^2 L^2}\right)^{5/6} {}_1F_1\left(-\frac{5}{6},1;\frac{2\rho^2}{W^2}\right)\right] \qquad \text{(B2)}$$

$$\langle\chi\rangle = -2.176 k^{7/6} C_n^2 L^{11/6} \text{Re}\left\{\left(\frac{6}{11}\right)i^{5/6}{}_2F_1\left(-\frac{5}{6},\frac{11}{6},\frac{17}{6};\frac{i\alpha L}{1+i\alpha L}\right)\right\} \qquad \text{(B3)}$$

$$\Delta_{\chi S} = 2.176 k^2 C_n^2 \int_0^L \left(\frac{(L-x)\gamma_I}{k}\right)^{5/6} \text{Im}\left\{{}_1F_1\left(-\frac{5}{6},1;-\frac{kP_{21}^2}{4(L-x)\gamma_I}\right) - {}_1F_1\left(-\frac{5}{6},1;-\frac{kP_{12}^2}{4(L-x)\gamma_I}\right)\right\} dx$$
(B4)

**ACKNOWLEDGEMENT**

The author would like to thank Dr. Albert Wheelon for his continuing support and inspiration during this work.

Collimated Beam Wave

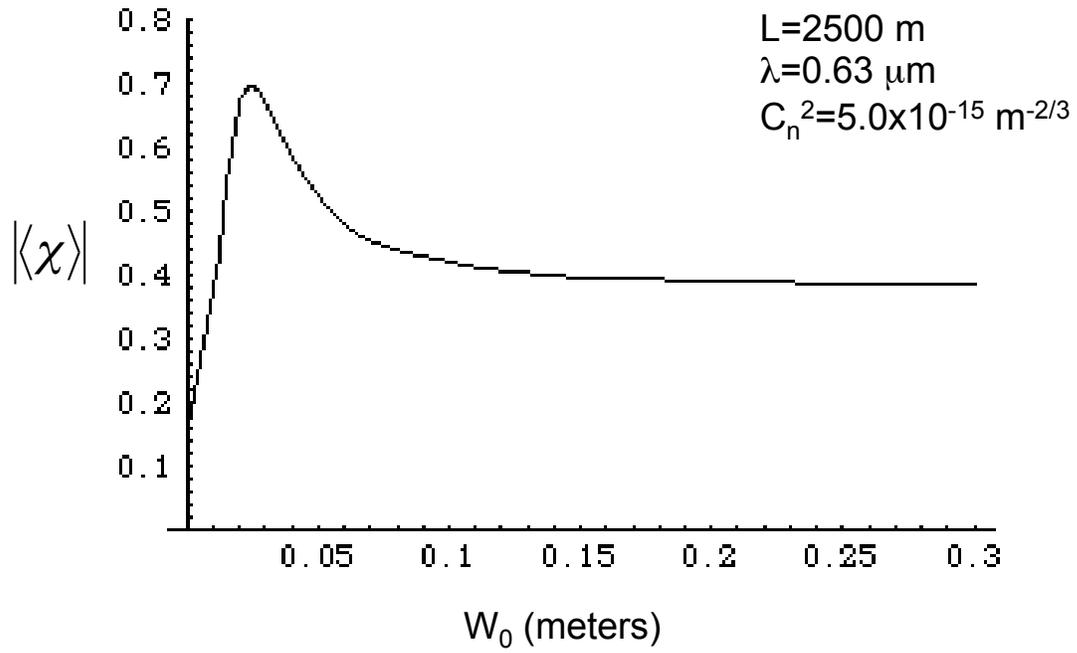

**Figure 1. Average Log-Amplitude for a Collimated Beam Wave**

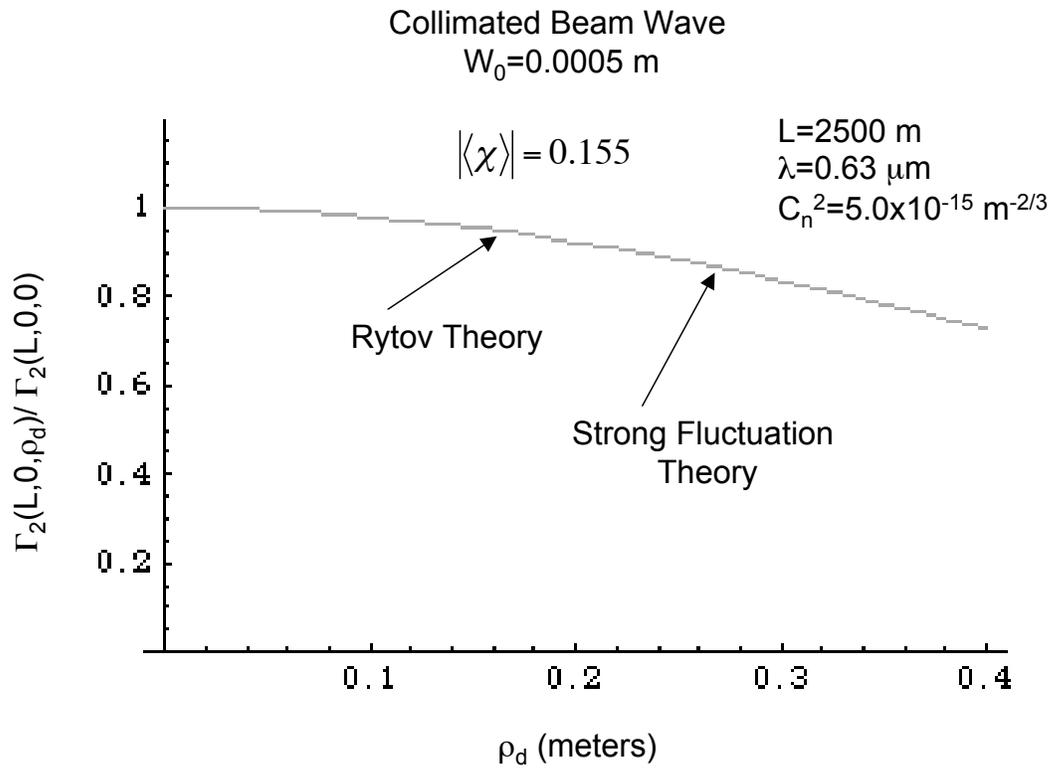

**Figure 2. Normalized Intensity for a Collimated Beam Wave of Waist Size=0.0005 m**

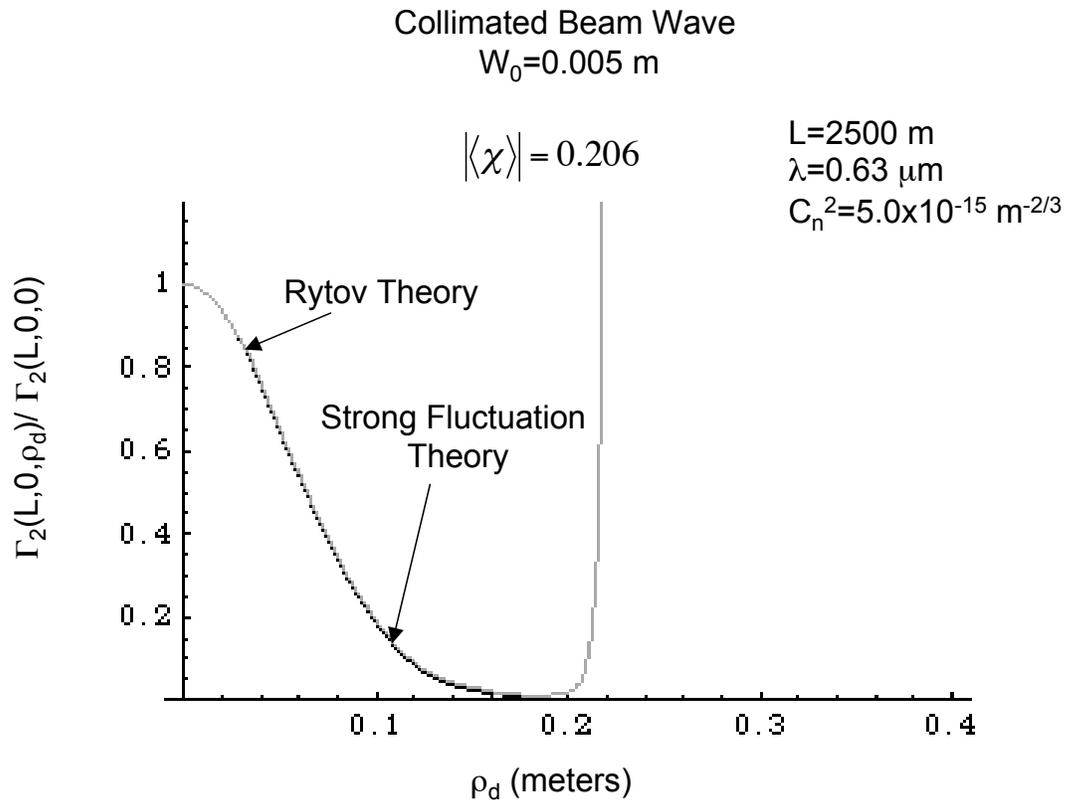

**Figure 3. Normalized Intensity for a Collimated Beam Wave of Waist Size=0.005 m**

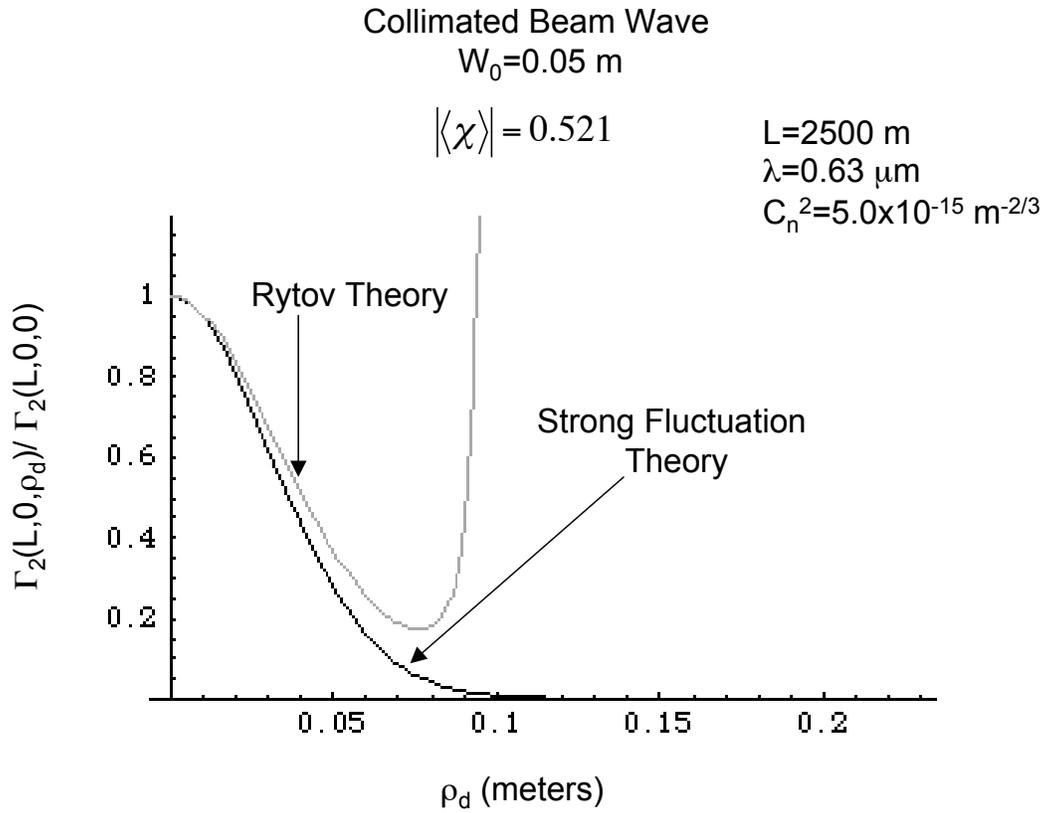

**Figure 4. Normalized Intensity for a Collimated Beam Wave of Waist Size=0.05 m**

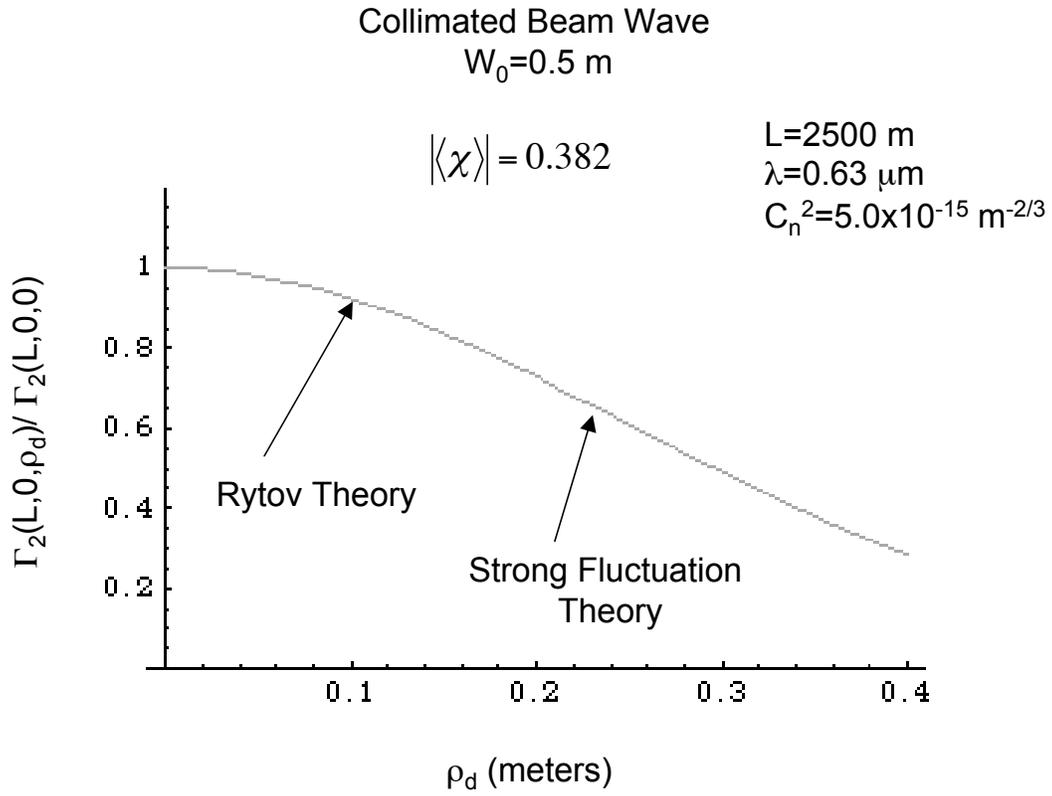

**Figure 5. Normalized Intensity for a Collimated Beam Wave of Waist Size=0.5 m**

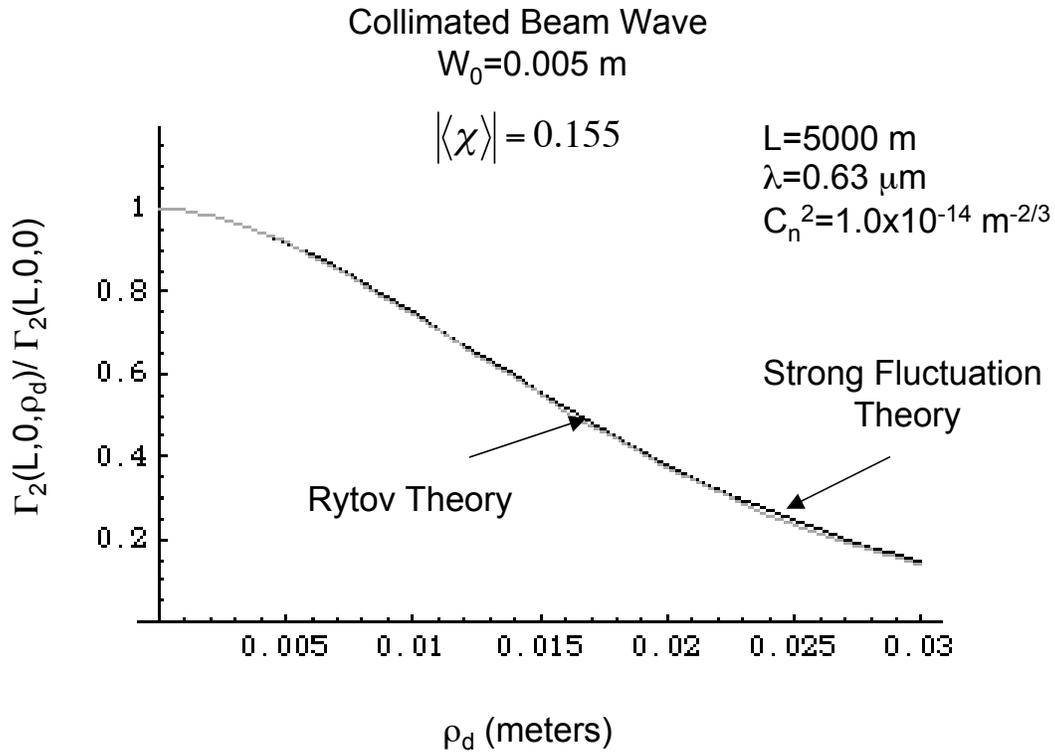

**Figure 6. Normalized MCF for a Collimated Beam Wave of Waist Size=0.005 m**

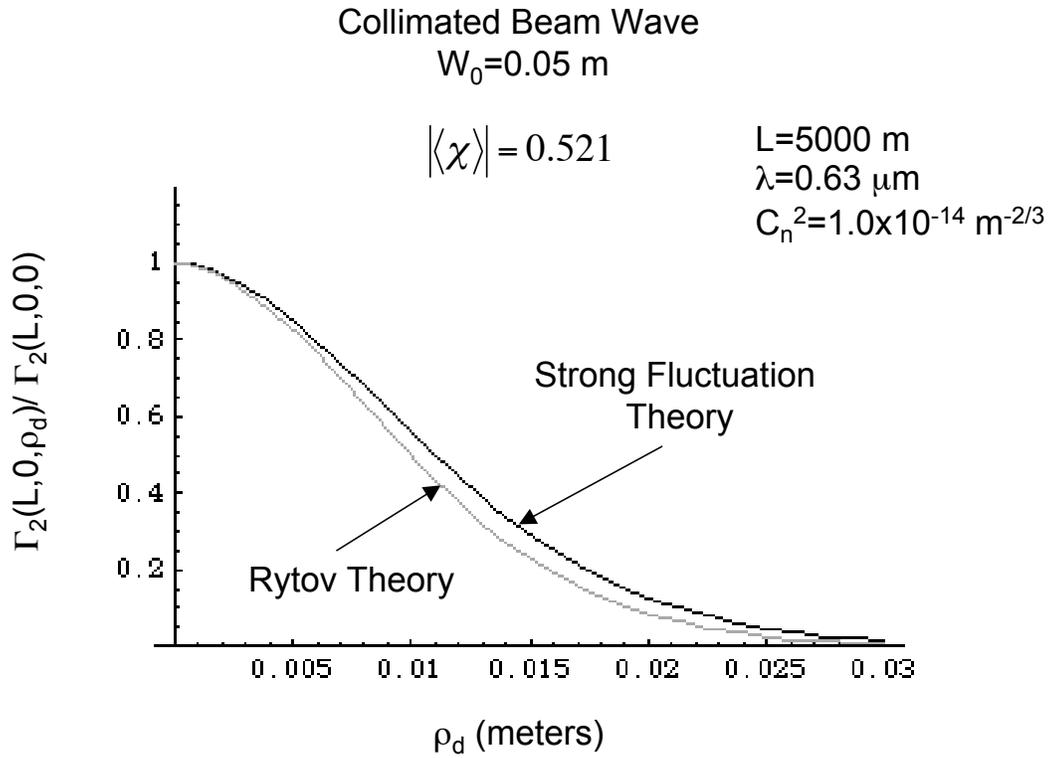

**Figure 7. Normalized MCF for a Collimated Beam Wave of Waist Size=0.05 m**

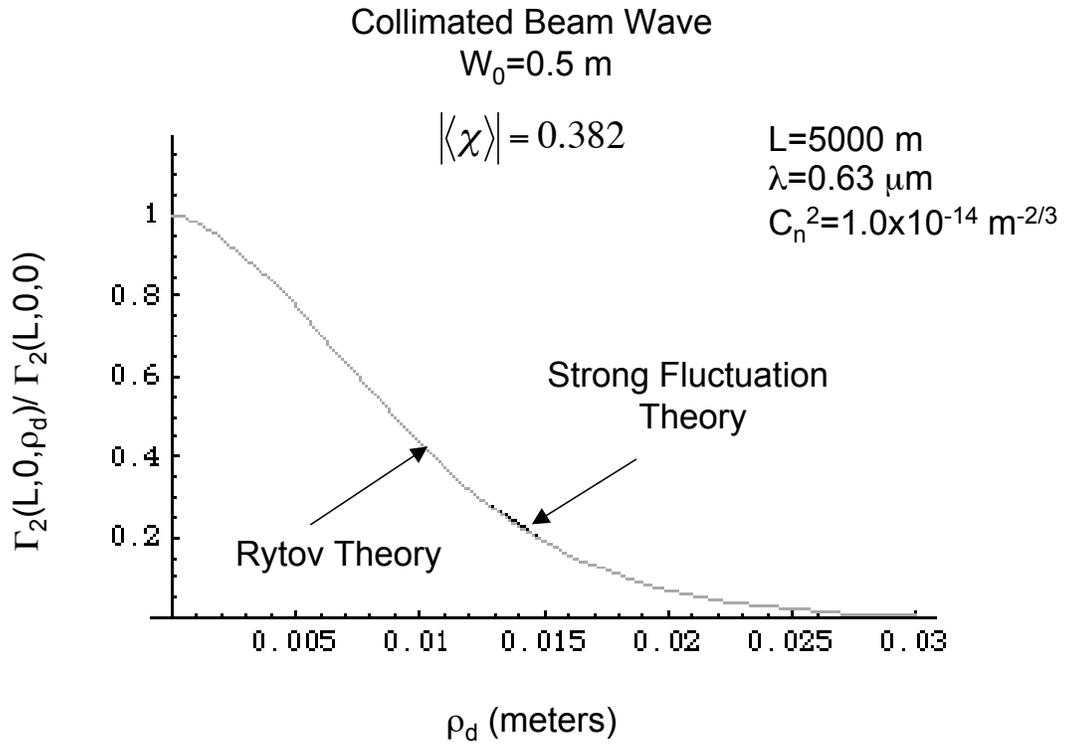

**Figure 8. Normalized MCF for a Collimated Beam Wave of Waist Size=0.5 m**